\documentclass{article}

\usepackage[utf8]{inputenc} 
\usepackage[T1]{fontenc}    
\usepackage{hyperref}       
\usepackage{url}            
\usepackage{booktabs}       
\usepackage{amsfonts}       
\usepackage{nicefrac}       
\usepackage{microtype}      
\usepackage{lipsum}
\usepackage{graphicx}
\graphicspath{ {./images/} }

\newcommand{\absdiv}[1]{%
  \par\addvspace{.5\baselineskip}
  \noindent\textbf{#1}\quad\ignorespaces
}

\title{Personalized acute stress classification from physiological signals with neural processes}

\author{
 Callum L Stewart$^1$ \\
  \texttt{callum.stewart@kcl.ac.uk} \\
\and
 Amos Folarin$^{1,2}$ \\
  \texttt{amos.folarin@kcl.ac.uk} \\
\and
Richard Dobson$^{1,2}$ \\
  \texttt{richard.j.dobson@kcl.ac.uk } \\
}

\date{
$^1$ Department of Biostatistics and Health Informatics, Institute of Psychiatry, Psychology and Neuroscience, King’s College London, London, U.K. \\
$^2$ Institute of Health Informatics, University College London, London, U.K. \\[2ex]
\today
}

\begin{document}
\maketitle
\begin{abstract}
\absdiv{Objective}
A person's affective state has known relationships to physiological processes which can be measured by wearable sensors. However, while there are general trends those relationships can be person-specific. This work proposes using neural processes as a way to address individual differences.

\absdiv{Methods}
Stress classifiers built from classic machine learning models and from neural processes are compared on two datasets using leave-one-participant-out cross-validation. The neural processes models are contextualized on data from a brief period of a particular person's recording.

\absdiv{Results}
The neural processes models outperformed the standard machine learning models, and had the best performance when using periods of stress and baseline as context. Contextual points chosen from other participants led to lower performance.

\absdiv{Conclusion}
Neural processes can learn to adapt to person-specific physiological sensor data. There are a wide range of affective and medical applications for which this model could prove useful.

\end{abstract}

\section{Introduction}
Wearable devices are increasingly used in affective computing because they provide continuous information on a person without requiring their attention; furthermore, some disorders or affective states have known relationships to measurable physiological processes \cite{10.1016/j.bspc.2015.02.012, 10.1007/978-3-540-89208-3_324} and are therefore candidates for remote monitoring. However, symptoms of disease and manifestations of affect can differ from one person to another, hindering the generalizability of models within mobile health and affective computing.

\subsection{Stress background}
Stress is a natural collection of responses to a change in homeostasis or a perceived threat. Stressful stimuli can elicit a variety of different behavioral, emotional, and physiological responses. The physiological response is predominantly mediated by the hypothalamic-pituitary-adrenal (HPA) axis and the autonomic nervous system (ANS) \cite{10.1038/nrn2647}, which in turn affect a range of physiological functions. In particular, the correlation between stress and heart rate and galvanic skin response (GSR) has long been known \cite{10.1097/00006842-196301000-00004}. Recent developments of wearable physiological sensors provide the ability for continuous, long-term, passive, and remote measurement. Their use, therefore, allows for an objective measure of systems mediated by the ANS and HPA in response to acute stress.

Accurate detection of stress has wide-ranging application. It could be used in intelligent feedback systems, altering the system’s behavior in response to stress \cite{10.1109/JSEN.2016.2533266}; as part of a system monitoring the progression of diseases or disorders with a known relationship with stress, such as depression \cite{10.1146/annurev.clinpsy.1.102803.144141}; detecting and managing maladaptive stress \cite{10.1109/ICDMW.2011.178}; or measuring response to medication or therapy.

\subsection{Machine learning personalization strategies}
Model personalization acknowledges that a single medical problem can have a diverse range of outcomes and symptoms between individuals, and attempt to improve model performance by making it specific to an individual. This has been approached in a number of ways; Non-exhaustively they include training completely separate models for each individual, selecting different features, setting personalized cut-off thresholds, additional training or hyperparameter selection of a general model using an individual’s data, and clustering individuals and creating a model for each cluster.

There are problems with some of the traditional personalization methods. Most generally, a lot of relevant data can be ignored if only a subset of a cohort is used to build a model, and the collection of adequate person-specific data is often time-consuming and expensive.  Few-shot and meta-learning techniques developed in adjacent fields, such as image classification, offer potential frameworks to approach the problem of personalizing models.

\subsection{Meta-learning and related approaches in biomedical datasets}
Various few-shot and meta-learning techniques have been developed recently. They are typically first used in generic open access few-shot datasets but have found some application in low-data domains like medical imaging.

Non-parametric or metric based networks, such as siamese networks and matching networks \cite{1606.04080}, learn a distance metric or comparable embeddings between input vectors. This method of few-shot deep learning appears to have had the greatest uptake in biomedical problems, with applications in seizure detection, histopathology \cite{10.1109/ISBI.2019.8759182}, drug discovery \cite{10.1021/acscentsci.6b00367}, and fall detection \cite{10.1145/3243250.3243268}, among others.

Optimization-based meta-learning, exemplified by MAML \cite{1703.03400}, aim to learn a set of initial parameters which can be quickly adapted to a new dataset through few additional gradient steps. They have found some promising use in low-data medical image classification tasks \cite{10.1007/978-3-030-00928-1_62, 10.1007/978-3-030-32239-7_17}, but are typically not used for personalizing a model to an individual in a longitudinal dataset.

Another technique, broadly categorized as parameterizing or black-box meta-learning, consists of a classification network and an encoder network which is used to parameterize the classifier. An example is neural processes \cite{1807.01622}. Neural processes are used as personalizable models in this study. They can be thought of as a distribution of functions, parameterised by a few ‘context’ x-y pairs. If we consider there to be an underlying biological trend to affective states or disorders, which manifest differently depending on the individual and their context, then a neural process forms a distribution of functions over the general trend which can be parameterized for a person using a small number of x-y pairs from that individual. Once trained on a meta-training set, individualization is provided only at the cost of a forward pass through the encoder, the original training data is no longer required.

\subsection{Study datasets and objective}
This work uses meta-learning techniques developed for few-shot learning to individualize models used for classifying periods of stress in participants from two publicly available datasets: Stress Recognition in Automobile Drivers (drivedb) \cite{10.1109/TITS.2005.848368} available from physiobank \cite{10.1161/01.cir.101.23.e215} and Wearable Stress and Affect Detection (WESAD) \cite{10.1145/3242969.3242985}. Both datasets contain continuous electrocardiogram (ECG) and galvanic skin response (GSR) recordings in healthy participants during a series of tasks designed to elicit an affective response. WESAD is a public dataset for affect and stress detection using motion and physiological recordings, including ECG and GSR. In addition to a task inducing stress, it includes an amusement task which is negatively labeled for our binary stress classification models. Drivedb is a dataset with multiple sensor recordings, including ECG and GSR, taken while a healthy participant drives on a predefined route containing sections of in-city driving assumed to be stressful, and sections of highway driving assumed to be relatively less stressful.

We investigate the applicability of neural processes (NPs), as a representative of meta-learning techniques, to personalization in a biomedical problem. Firstly, the performance of general models built with k-Nearest Neighbors (k-NN), support vector machine (SVM) with a radial basis function kernel, and an L1-regularized logistic regression (Lasso), are compared against a neural process individualized with either baseline-only or randomly chosen context pairs. Secondly, a baseline-only personalization is compared against a neural process which uses the first highway and city driving segments in addition to the baseline in the drivedb dataset, which includes repeated sections intended to induce stress (city) and relatively reduced stress (highway). The sections used as context points are excluded from the test performance.

\section{Methods}
\begin{figure}[t]
  \centering
    \makebox[\textwidth][c]{\includegraphics[width=1.2\textwidth]{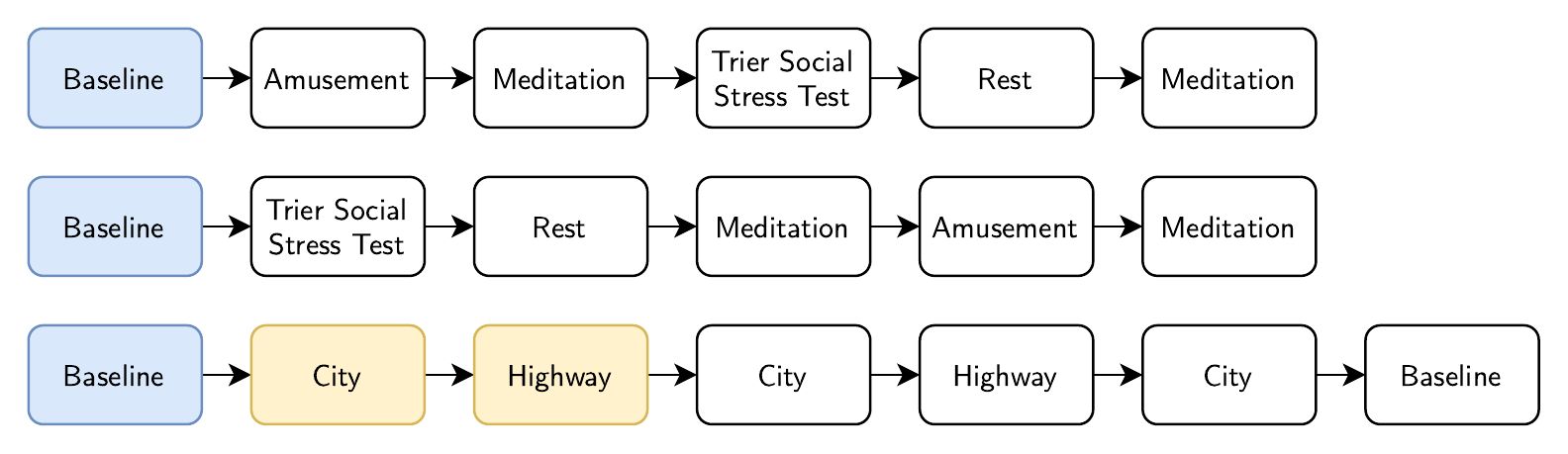}}
  \caption{Study protocols for datasets. The two protocol variations in WESAD (top, middle) and the route driven in drivedb (bottom).}
  \label{fig:study_protocols}
\end{figure}

The WESAD dataset contains 15 participants (age $= 27.5 \pm 2.4$) with an average recording duration of 96 minutes. Two affective responses, stress and amusement, are evoked in two tasks. There is a baseline period, rest period and two meditation tasks \ref{fig:study_protocols}. Two devices are worn by participants, but only ECG and GSR signals from the chest-worn RespiBAN are used here.
Drivedb contains recordings of 17 participants lasting between 65 and 93 minutes, depending on the road conditions during the test. Of these, only 13 are used (4-16) because of missing data or unclear label markers. Again, only the ECG and GSR signals are used, collected from a custom wearable system.

\subsection{Processing and feature extraction}
Manually defined features are used to facilitate comparison between general machine learning models and the neural process models, and to evaluate whether the neural process is able to use data representative of an individual to improve performance rather than the ability of a neural network to learn representations from raw or preprocessed signals. 

The Hamilton-Tompkins algorithm is used to detect the R peaks in the ECG signal \cite{10.1109/TBME.1986.325695}. Tonic and phasic GSR are filtered from the raw GSR signal using a 0.2Hz lowpass and a 0.5-2Hz bandpass filter respectively; both are 5th order Butterworth filters. Participants in the drivedb dataset can contain GSR recorded at either the hand, foot, or both. Only one is used, and the hand GSR is preferred if it is available.

Typical heart rate, heart rate variability (HRV), phasic GSR, and tonic GSR features are extracted in 40s windows with 20s overlap from each participant \ref{tab:features}. The class label for the window is determined by the largest proportion of time spent in either the stressful or relaxing task. Features are min-max scaled between -1 and 1 for all general models, but are not scaled for the neural processes.

\begin{table}
    \centering
    \begin{tabular}{  l  l  l }
        Signal & Feature & Equation / Reference \\
        \hline
        ECG & HR range & $max(HR) - min(HR)$\\
        ECG & HR mean & $\bar{x} = \frac{1}{N}\sum_{i=1}^N x_i$\\
        ECG & HRV SDNN & \cite{10.3389/fpubh.2017.00258} \\
        ECG & HRV RMSSD & \cite{10.3389/fpubh.2017.00258} \\
        ECG & HRV CSI & \cite{10.1109/EMBC.2014.6944639} \\
        ECG & HRV sample entropy & \cite{10.1152/ajpheart.2000.278.6.H2039} \\
        ECG & RQA determinism & \cite{10.1016/j.physrep.2006.11.001} \\
        ECG & RQA length entropy & \cite{10.1016/j.physrep.2006.11.001} \\
        ECG & HRV LF absolute power & \cite{10.3389/fpubh.2017.00258} \\
        ECG & HRV LF relative power & \cite{10.3389/fpubh.2017.00258} \\
        ECG & HRV LF peak frequency & \cite{10.3389/fpubh.2017.00258} \\  
        ECG & HRV HF absolute power & \cite{10.3389/fpubh.2017.00258} \\
        ECG & HRV HF relative power & \cite{10.3389/fpubh.2017.00258} \\
        ECG & HRV HF peak frequency & \cite{10.3389/fpubh.2017.00258} \\
        ECG & HRV HF/LF ratio & \cite{10.3389/fpubh.2017.00258} \\
        GSR tonic & Mean & $\frac{1}{N}\sum_{i=1}^N x_i$ \\
        GSR tonic & SD & $\sqrt{\frac{1}{N-1} \sum_{i=1}^N (x_i - \bar{x})^2}$\\
        GSR tonic & 1st deriv. mean & $\frac{1}{N}\sum_{i=1}^N x'_i$\\
        GSR tonic & 1st deriv. SD &  $\sqrt{\frac{1}{N-1} \sum_{i=1}^N (x`_i - \bar{x`})^2}$ \\
        GSR phasic & SD & $\sqrt{\frac{1}{N-1} \sum_{i=1}^N (x_i - \bar{x})^2}$\\
        GSR phasic & mean absolute value & $|\bar{x}| = \frac{1}{N} \sum_{i=1}^N |x_i|$\\
    
    \end{tabular}
    \label{tab:features}
    \caption{Feature definitions}
\end{table}

\subsection{General model}
Both general and personalized models are trained and tested using leave-one-participant-out cross-validation. Traditional machine learning models were built using scikit-learn \cite{1201.0490}, using default hyperparameters. Three general models are used: A logistic regression with l1 penalization and $C=1.0$, a radial basis function kernel SVM with gamma = 0.0526 ($1 / N_{features}$), and a 20-neighbor k-NN classifier. 

\subsection{Personalized neural process models}
\begin{figure}
  \centering
    \makebox[\textwidth][c]{\includegraphics{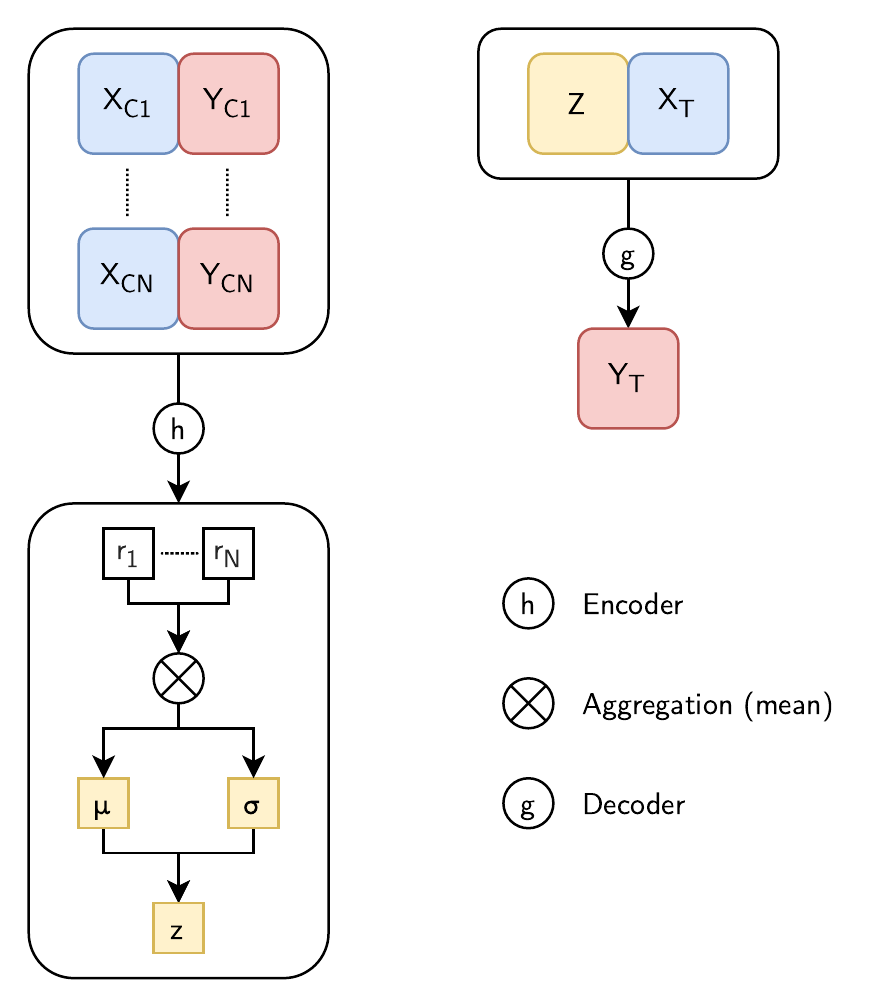}}
  \caption{A general view of a neural process model. The encoder (h) takes X-y pairs and transforms them into a latent distribution. A sample from that distribution is concatenated with unlabeled data and passed to the decoder (g) which predicts class labels for the unlabeled data.}
  \label{fig:np_generic}
\end{figure}

A neural process is a latent variable neural network which aims to adapt to specific context data at test-time. It is formed of three components: an encoder $h(r_i | x_{ci}, y_{ci})$ which transforms a set of contextual input data pairs $(x_c, y_c)$ into a representation $r$, an aggregator which aggregates multiple representations from the encoder into a single vector which is used to parameterize a latent distribution $z$, and a decoder $g(y_t | x_t, z)$ which samples z and transforms unlabelled data $x_t$ into a predicted value $y_t$ \ref{fig:np_generic}. The model is built using the pytorch library \cite{1912.01703}.

The specific architecture used here consists of an encoder of 3 dense hidden layers each with 30 nodes, a mean aggregator and a latent variable with 15 nodes, and a decoder with 3 dense 30 node hidden layers and a single output node. The decoder has a dropout rate of 0.2. Each model is trained on the data of all but one participant.

\begin{figure}
  \centering
    \makebox[\textwidth][c]{\includegraphics[width=1.2\textwidth]{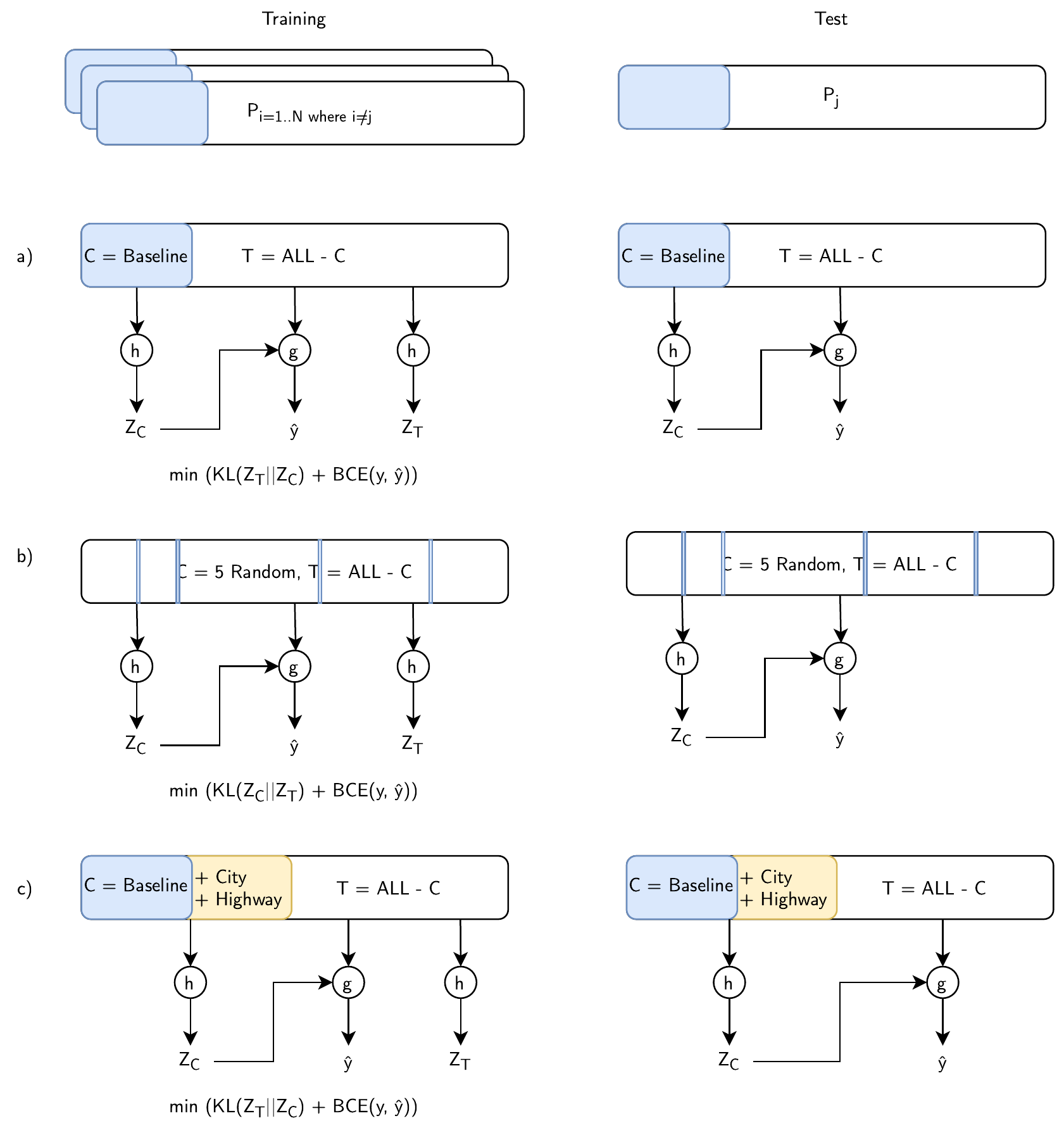}}
  \caption{Context strategies for training and testing in each participant’s (j) neural process model. h: Encoder, g: Decoder, BCE: Binary cross entropy, KL: Kullback–Leibler divergence. In each case, the training loss is the sum of the Kullback-Leibler divergence between the two distributions formed by target and context points passing through the encoder and the log-loss between y and y-pred. a) Context points are taken from the baseline recording, target points are taken from the remaining data. b) Context points are randomly selected using a uniform distribution. c) Drivedb only - Context points are taken from the baseline, first city, and first highway sections. An equal number of points are taken from each section.}
  \label{fig:training_strategies}
\end{figure}

During training, the data of each training participant is looped through. Between 5 and 10 points are randomly chosen as the context points. All of the data belonging to the current training participant is used as target points. Both context ($xy_c$) and target ($xy_t$) x-y pairs are passed through the encoder to create latent distributions $Z_c$ and $Z_t$ respectively. The encoded context points ($Z_c$) are concatenated with the unlabeled target points ($x_t$) and passed through the decoder to predict the target label ($\hat{y}_t$). As well as minimizing the binary cross-entropy between the predicted target points ($\hat{y}_t$) and the true values ($y_t$), the Kullback Leibler divergence between the distributions of the encoded context points ($Z_c$) and the encoded target points ($Z_t$) is minimized (\ref{fig:training_strategies}).

At test time the context points are unique from the target points and selected according to the personalization strategies mentioned in the following paragraph. In each case 6 x-y data pairs are used as the context points. Any period or task from which the context points are chosen are not used as target points. Stress predictions for the remaining data for the participant are calculated.

\subsection{Personalization strategies}
Three methods for selecting context points during testing are chosen. Firstly, each model is personalized using context points selected only from the baseline segment. Secondly, context points are randomly selected from the entire recording with a uniform distribution. Thirdly, two points from each of the baseline recording and the first occurrence of the city (stress) and highway (non-stress) driving segments are used as context points. Data from the recording following the sections chosen for context points; two city driving sections, a highway section, and a relaxation section; are subsequently predicted using the personalized model. Because the WESAD dataset only has a single stress assessment task, only the drivedb dataset can be used in the third personalization strategy.

\subsection{Performance metrics}
Three performance metrics are reported here: the area under the curve (AUC) of the receiver operating characteristic (ROC), the average precision, and the log-loss. Because of the class imbalance in the WESAD dataset and the differences in class proportion between datasets, the average precision may be more informative than the AUC. 

\section{Results}
Both the baseline-only and randomly chosen context NPs perform better than all of the general models, with the randomly chosen context performing best \ref{tab:results_wesad}. Randomly chosen points, while not included in the test scores themselves, are likely to be strongly correlated with points from the surrounding time and from the same task. The third personalization strategy, in which the context points are selected from the baseline and  first occurrence of each task, is used to address this problem in models for drivedb participants. Similarly to the previous results, the personalized NP models perform best and the models which include stress-task context perform better than those which use baseline-only data \ref{tab:results_drivedb}, although the improvement in comparison to the baseline-only models is much slighter.

\begin{table}
    \centering
    \begin{tabular}{l l l l}
    Model &
    AUC &
    Average precision &
    Log loss \\
    \hline
    Lasso &
    0.954 &
    0.881 &
    0.222 \\
    SVC (RBF kernel) &
    0.943 &
    0.882 &
    0.234 \\
    K-Nearest Neighbors &
    0.870 &
    0.740 &
    0.563 \\
    NP (baseline) &
    0.970 &
    0.924 &
    0.182 \\
    NP (random choice) &
    0.984 &
    0.957 &
    0.133 \\
    NP (other participant) &
    0.880 &
    0.780 &
    0.470 \\
    \end{tabular}
    \label{tab:results_wesad}
    \caption{WESAD dataset results}
\end{table}

\begin{table}[]
    \centering
    \begin{tabular}{l l l l}
Model &
AUC &
Average precision &
Log loss \\
\hline
Lasso &
0.695 &
0.736 &
0.669 \\
SVC (RBF kernel) &
0.704 &
0.733 &
0.645 \\
K-Nearest Neighbors &
0.680 &
0.721 &
3.09 \\
NP (baseline) &
0.776 &
0.797 &
0.570 \\
NP (tasks) &
0.787 &
0.804 &
0.553\\
NP (other participant) &
0.722 &
0.757 &
0.663 \\

    \end{tabular}
    \caption{Drivedb dataset results}
    \label{tab:results_drivedb}
\end{table}

\begin{figure}
  \centering
    \includegraphics[width=\textwidth]{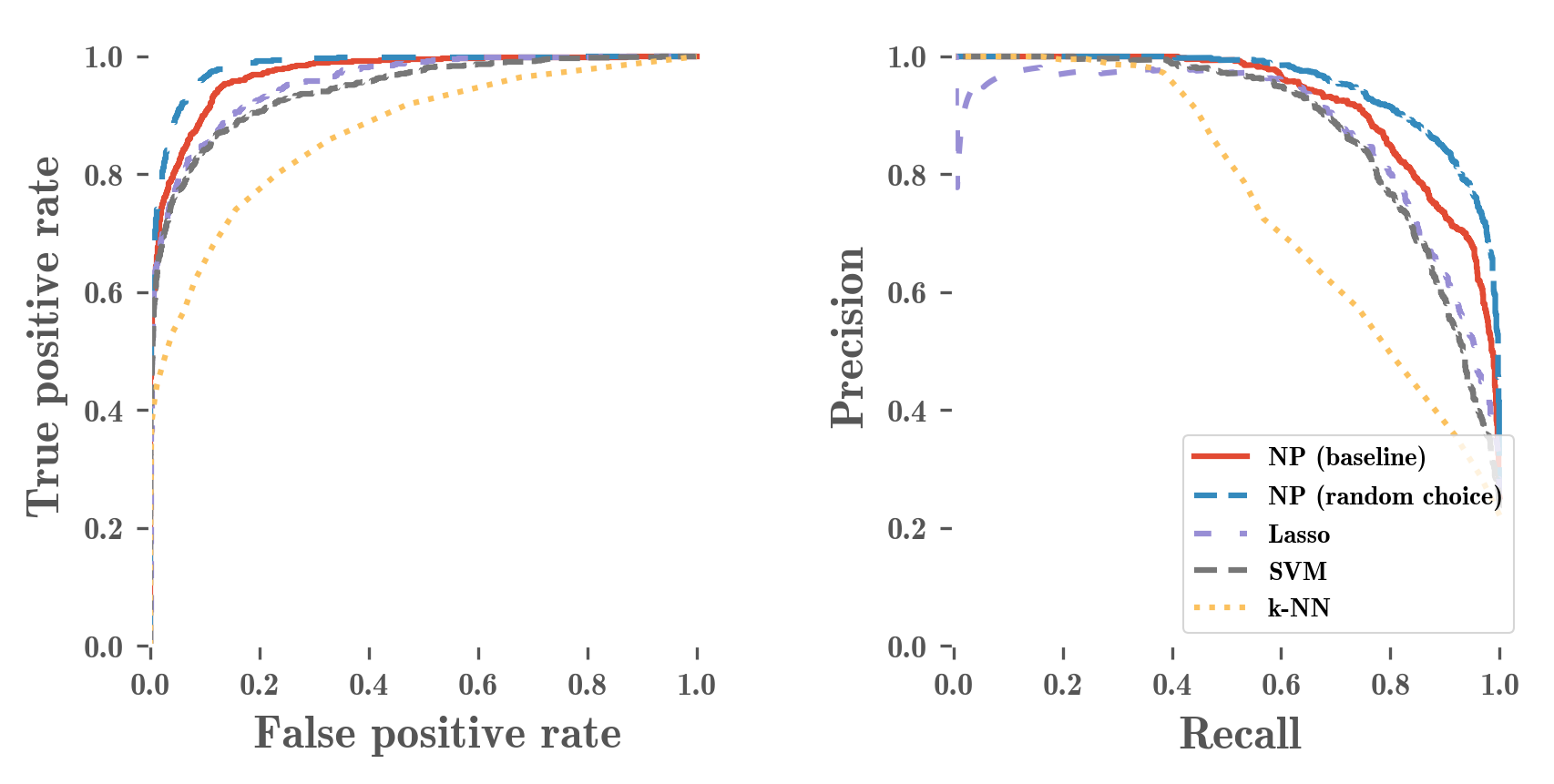}
    \includegraphics[width=\textwidth]{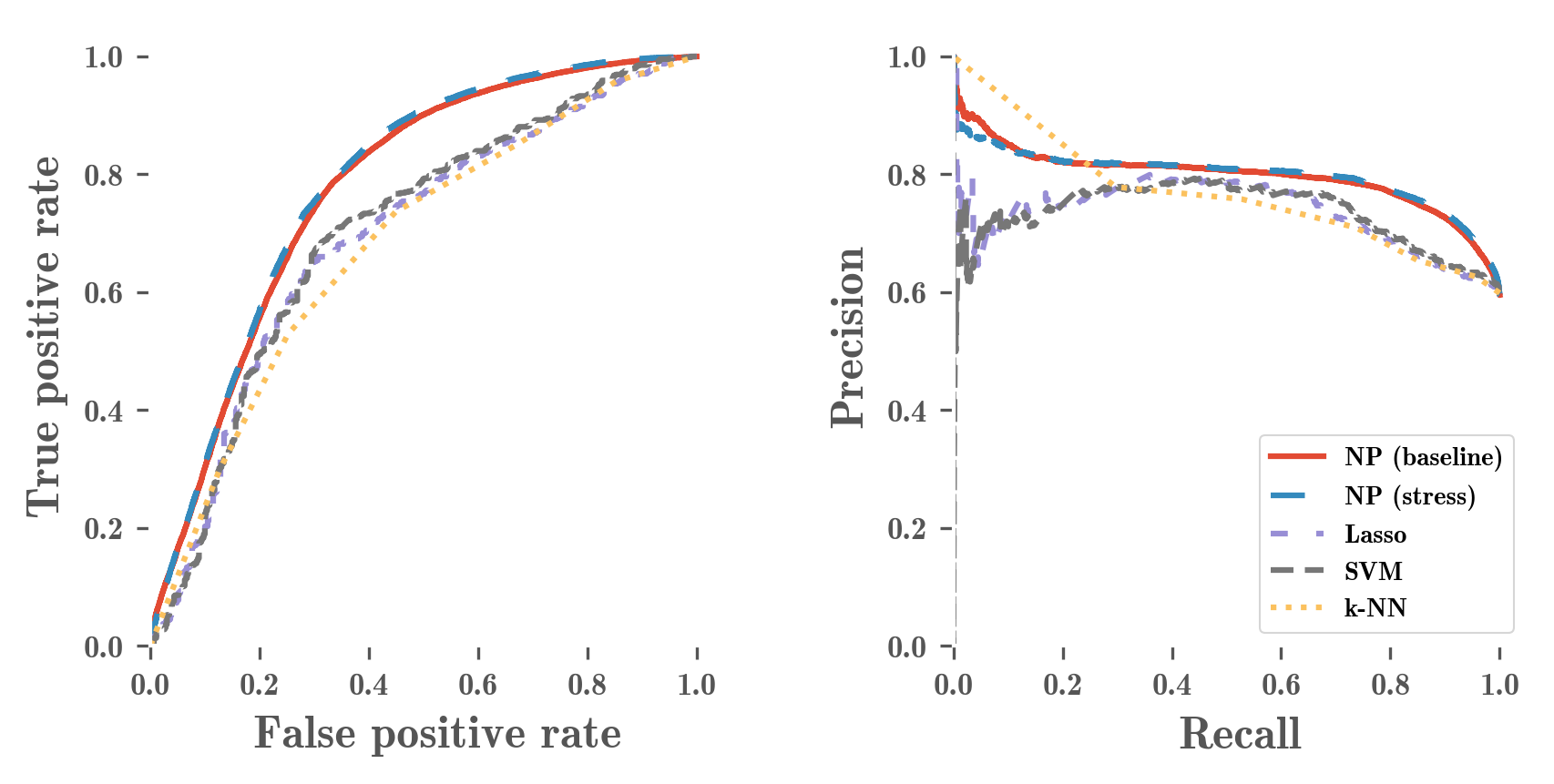}
  \caption{Receiver operating characteristic (left) and precision-recall (right) plots for WESAD (top) and drivedb (bottom) models.}
  \label{fig:model_comparison}
\end{figure}

\begin{figure}
  \centering
    \includegraphics[width=\textwidth]{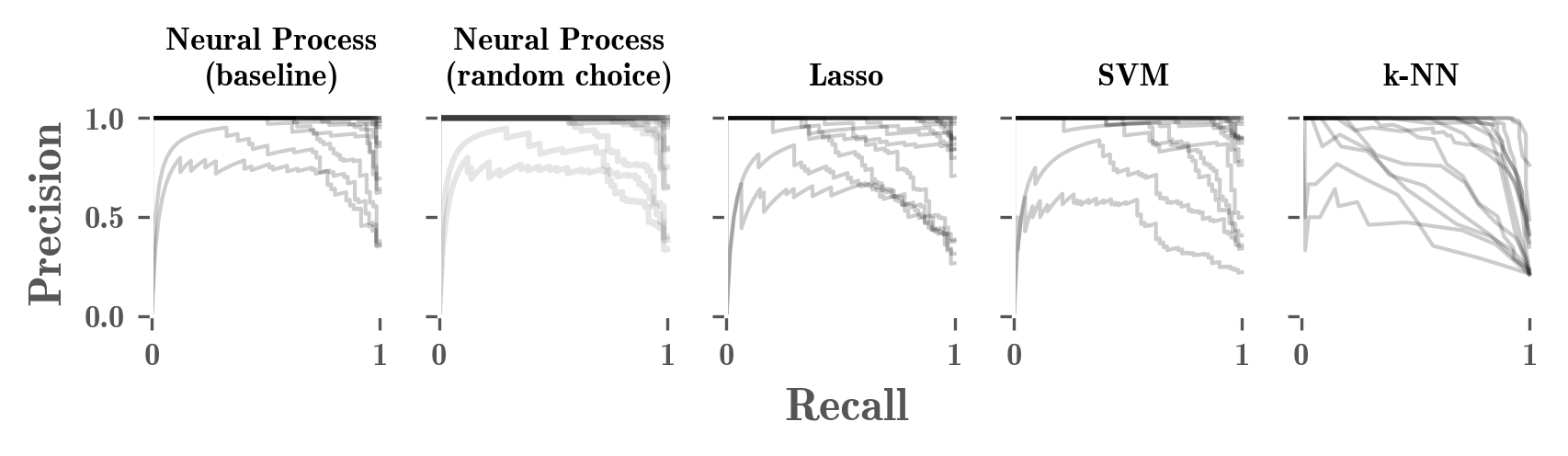}
  \caption{Precision-recall plots for each model in which models belonging to participants from the WESAD dataset are drawn as individual lines.}
  \label{fig:indiv_results}
\end{figure}

Using the neural process models with context and target points selected from different participant results in greatly reduced performance (WESAD: 0.957 average precision, same-participant vs 0.780 other-participant, drivedb: 0.804 vs 0.757), indicating that the neural processes do rely on and gain benefit from individual-specific data points \ref{fig:model_comparison}.
The increase in performance between the general and personalized models appears partly due to a decrease in variance of performance between each participant’s model \ref{fig:indiv_results}. Each of the general models have a subset of low-performing participants, although they do not completely overlap. The distribution of performance over participants for the general models contains a large number of participants with very high performance, 0.9+ accuracy, and a tail of drastically lower performing participants. The personalized NP models performance is mostly improved through increased performance on those lower performing participants. Within the WESAD cohort, the average accuracy between the NP and lasso models was increased by 0.0995 for those participants whose accuracy was less than 0.9 in the general lasso model, compared to a decrease of 0.002 for those above 0.9 in the lasso model. The drivedb dataset shows a broadly similar pattern; participants with an accuracy score in the general lasso model lower than the average across the dataset (0.66) have an average accuracy increase of 0.117, compared to an increase of 0.035 in those who had lasso model accuracies above the average.

\section{Discussion}
Overall the personalized models performed better than the general models. Encouragingly, the improvement between personalized and general models is most marked in those participants with lower performance in the general models. The improved performance of the lower performing subgroup in the NPs suggests that by using a small amount of person-specific data, a base model can be successfully used in a greater range of people who are dissimilar to the training cohort or who are somehow divergent in comparison to the majority. The demographics for the participants in the two studies was quite homogenous; given a broader population, the usefulness of a more flexible or personalizable model may be greater.

There are two remaining participants with low-performing personalized models, visibile in \ref{fig:indiv_results} and both belonging to the WESAD cohort. One has low performance across all models, where no model can differentiate the amusement and stress task well. The second is particular to the NP model, and appears to be due to an atypical baseline recording, which includes a large tonic GSR amplitude, highlighting the importance of contextual data that is  typical of the class it is representing. If the first few minutes of the recording are discarded, and a portion of the first rest period selected as the context, the performance of the model is similar to other participants.

Model personalization methods previously used in stress detection studies have been typically achieved through personal feature normalization \cite{10.1109/FG.2015.7284844}, training a model on one participant’s data\cite{10.1109/jbhi.2015.2446195, 10.1007/978-3-642-24600-5_16, 10.1145/2370216.2370270, 10.5220/0007368802560263}, or training models for groups of similar participants \cite{10.1109/jbhi.2015.2446195, 10.1109/TAFFC.2016.2610975}. Neural processes have a number of theoretical advantages over these methods; they do not assume that personal differences in features can be reduced to a linear proportion of a baseline measurement, they can make use of the entire dataset of participants, only a small number of labelled data points are required to personalize a model, and the computational cost of personalization is only a single pass through the encoder network.

The importance of correct use of cross-validation or training splits is demonstrated in the literature. High performance can be achieved when an individual model is trained using random cross-validation \cite{10.1109/TAFFC.2016.2610975, 10.5220/0007368802560263} because temporally close data points will be highly correlated. This is also seen in the neural process models, in which randomly sampled context data points outperform context from a single task. For the purpose of building personalized models, it is therefore necessary to have a dataset with multiple assessments per participant, or to personalize based on unlabeled or negative case data.

In general, using meta learning techniques additional medical datasets with similar tasks and signals could be incorporated. Aggregation of similar small datasets could lead to improved performance for each individual task. To combine multiple tasks along with personalization through meta-learning, it may be necessary to pose the meta-learning procedure in multiple levels or hierarchically \cite{1905.05301v1}, where more prior knowledge is shared between individuals in the same task than between the different tasks.

In this study features are manually defined and extracted from the raw signals because the objective was to discover whether personalization through neural processes is possible and useful, rather than looking at whether a neural network can learn a better feature representation. However, meta-learning can allow more sophisticated deep learning techniques and feature extraction where they would otherwise be intractable because of small dataset sizes. Addition of a neural network to learn features is therefore a prominent area to potentially improve performance.

That only baseline data points used as context can improve performance suggests that a representation built on unlabeled or weakly labeled data may be viable. Particularly in long-duration recordings, much of the data in biomedical datasets can be unlabeled or with a very large imbalance between positive and negative cases. Going forward, it would therefore be useful to be able to personalize a model based on that unlabeled data. Where the representations from the context data points are currently mean aggregated, it may make more sense to have an aggregation that recognizes the time-dependent nature of the data. Additionally, in the future it would be useful to compare the performance of different personalization techniques, both optimization-based deep learning and classical machine learning, against the neural processes demonstrated here.

\section{Conclusion}
Neural processes, presented as a method to generate personalized models, outperform general classic machine learning algorithms in stress detection tasks across two datasets and appear to use small amounts of person-specific context data to improve performance. Using only baseline data as context is useful, but the inclusion of data with the positive-label class further improves performance. The datasets used here concern affect and stress classification, but there are applications beyond: many problems in medicine and biology have large inter-individual differences or heterogeneity in classification which could be addressed using neural processes or similar methods. There is also a large space for improvement in various aspects of the modeling procedure.

\section*{Acknowledgments}
This paper represents independent research funded by the National Institute for Health Research (NIHR) Biomedical Research Centre at South London and Maudsley NHS Foundation Trust and King’s College London. The views expressed are those of the authors and not necessarily those of the NHS, the NIHR or the Department of Health and Social Care.

\bibliographystyle{unsrt}  
\bibliography{references}
\end{document}